# Mechanical Properties of Pristine and Defective Carbon-Phosphide Monolayers: A Density Functional Tight-Binding Study

V. Sorkin* and Y.W. Zhang†

Institute of High Performance Computing, A*STAR, Singapore 138632

## Abstract

Using density functional tight-binding theory, we investigated the elastic properties and deformation and failure behaviors of pristine and defective carbon-phosphide (CP) monolayers subjected to uniform uniaxial tensile strain along armchair (AC) and zigzag (ZZ) directions. Two variants of carbon phosphide (α-CP and β-CP) were studied and two types of carbon and phosphorous vacancies (single and double) were considered. It was found that carbon monovacancies have the lowest formation energy, while phosphorous divacancies have the highest one in both CP allotropes. A strong mechanical anisotropy for carbon phosphide was found with the Young's modulus and the failure stress along ZZ direction being an order of magnitude larger than those along AC direction. In both allotropes, the Young's modulus, failure stress and strain are considerably affected by vacancies, especially along AC direction. Fracture of pristine CP monolayer occurred via the rupture of phosphorous-phosphorous bonds when CP monolayer is stretched along AC direction, while via the rupture of carbon-phosphorous bonds when stretched along ZZ direction. Defective α-CP and β-CP monolayers both undergo a brittle-like failure initiated around the hosted vacancies at a lower critical strain. The failure strain and stress along the AC direction are affected only by phosphorous vacancies, while along the ZZ direction, they are almost equally affected by both phosphorous and carbon vacancies. These understandings may provide useful guidelines for potential applications of CP monolayers in nanoelectromechanical systems.

## 1. Introduction

Two-dimensional (2D) materials with semiconducting properties are of prime interest for basic research and potential technological applications. Recently, a new semiconducting 2D material, named carbon phosphide (CP) (also called phosphorus carbide), with superior carrier mobility consisting of $sp^2$ hybridized carbon atoms and $sp^3$ hybridized phosphorous atoms, has been proposed [1–3]. This interesting material can exist in a few stable phases with puckered or buckled hexagonal lattice. Two of its allotropes, that is, α-CP and β-CP, are semiconductors with a finite band gap. Remarkably, their charge carrier mobilities is five times higher than those of phosphorene [2].

---

* Email address: sorkinv@ihpc.a-star.edu.sg
† Email address: zhangyw@ihpc.a-star.edu.sg

The feasible phases of carbon phosphide monolayer were first predicted by Wang et al. [1] by using the particle swarm optimization method coupled with density functional theory (DFT) calculations. Examination of the electronic structure of carbon phosphide indicated that α-CP monolayer is a semiconductor with an indirect band gap of $E_{bg}$=1.26 eV, while β-CP monolayer is a semiconductor with a direct band gap of $E_{bg}$=0.87 eV [1]. The electronic properties of α-CP and β-CP monolayers were found to be highly anisotropic: the effective mass of charge carriers along ZZ direction was significantly larger than that along AC direction. Correspondingly the carrier mobility in AC direction was found to be considerably larger than in ZZ direction. In addition, phonon dispersion calculations proved the stability of the predicted phases [1]. Singh et al. [2] investigated the electro-optical and thermoelectric properties of carbon phosphide monolayers. They found that large absorption of radiation of carbon phosphide in ultraviolet range and significant transmission in visible range can be potentially used in optical devices, particularly as an anti-reflecting layer in solar cells. The outstanding thermal properties also make carbon phosphide a promising thermoelectric material [2].

The list of the potential carbon phosphide phases was lately extended by Guan et al. [3], and a variety of carbon phosphide allotropes with metallic, semi-metallic and semiconducting properties were predicted. These different phases were found to arise from subtle competition between $sp^2$ bonding of carbon and $sp^3$ bonding of phosphorous atoms. Using first-principles calculations, Rajbanshi et al. [4] confirmed the stability of a few newly proposed phases. In addition, they also explored the effects of hydrogenation of carbon phosphide monolayers and homogeneous doping by phosphorous atoms in graphene and by carbon atoms in phosphorene. Zhang et al. [5] investigated the effect of molecular doping (by $O_2$, $H_2O$, $NO_2$, and $NH_3$ molecules) on the electronic and optical properties of carbon phosphide monolayers by using first-principles calculations. They found that the molecules act as electron acceptors and introduce deep acceptor energy levels in the band gap.

Several attempts have been made towards the experimental fabrication of carbon phosphide monolayers. For example, Jones et al. [6] synthesized bulk carbon phosphide by doping diamond-like carbon with phosphorous atoms. Kuo et al. [7] carried out the synthesis of amorphous carbon phosphide films using radio-frequency plasma chemical vapor deposition. Hart et al. [8] synthesized 3D bulk carbon phosphide using pulsed laser deposition. Hence, carbon phosphide monolayers can be conceivably exfoliated from these synthesized 3D bulk CP-samples [8] or can be potentially directly grown by chemical vapor deposition [7].

We note that previous theoretical studies were mostly concentrated on the electronic and thermal properties of CP monolayers. The mechanical properties, especially deformation and failure behaviors of CP monolayers, remain largely unexplored. Besides that, previous work was concentrated entirely on the properties of pristine carbon phosphide. However, the presence of defects, such as point and line defects, is generally inevitable. Since the knowledge of the mechanical properties of pristine and defective carbon phosphide is crucial to the operating reliability of CP monolayer-based nanoelectromechanical devices, hence, it is both important and necessary to gain such understanding in order to offer useful guiding principles for design, fabrication and potential applications of carbon phosphide.

Here we study the effects of vacancies on the mechanical properties of pristine and defective carbon phosphide, focusing on the elastic properties, tensile deformation and failure behavior. Two allotropes of carbon phosphide were investigated: α-CP and β-CP. Two types of carbon and phosphorous vacancies were introduced: single and double vacancies. Uniaxial tensile strain was applied along the two primary directions: arm-chair (AC) and zig-zag (ZZ). Essentially, we would like to have answers to the following questions: What are the atomistic structures around the hosted vacancies? How large are the vacancy formation energies? What are the Young's modulus and Poisson's ratio of pristine and defective carbon phosphide? How strong is the anisotropy in the mechanical properties? In what way the mechanical properties are affected by the hosted vacancies? How carbon phosphide deforms under uniaxial tensile strain? What is the failure mechanism of pristine and defective carbon phosphide? How the failure strain and failure stress depend on the type of the hosted vacancies and the direction of applied tensile strain? To resolve these questions, we carry out density functional tight binding simulations studying carbon phosphide monolayers subjected to uniaxial tensile strain.

## 2. Computational Model

We applied the tight-binding (TB) method [9–11] to investigate the deformation and failure of pristine and defective CP monolayers under uniaxial tensile strain. The TB method was chosen since the density function theory (DFT) is too computationally demanding for the moderately large CP-samples (up to ~500 atoms) containing vacancies, which are subjected to large tensile deformations resulting in their ultimate failure. Molecular dynamics (MD) simulations cannot be used since an interatomic potential for carbon phosphide is currently not available. In addition, MD simulations produce less accurate results since they disregard all the electronic degrees of freedom and related quantum effects. Therefore, the TB technique that provides an appropriate balance between MD efficiency and DFT precision is the ideal method to deal with the above-mentioned limitations.

In our simulations of pristine and defective carbon-phosphide layers, we applied the density functional tight-binding (DFTB) method [11,12]. This method was derived as a TB approximation of the DFT method. It employs a number of empirical approximations to increase its computational efficiency, while maintaining to a large extent the underlying DFT accuracy [11,13–15]. The replacement of the many-body Hamiltonian of DFT with a parameterized Hamiltonian matrix is the crucial approximation of the DFTB [11,16]. A few additional terms are included to the Hamiltonian matrix to model the short-range repulsion, van der Waals and Coulomb interactions [17]. Furthermore, the self-consistent charge (SCC) method is applied to further enhance the description of atom bonding in DFTB [17]. Lately, the DFTB method has been applied to examine the mechanical properties of pristine and defective graphene [18–20] and phosphorene monolayers, nanoribbons, nanotubes, and grain boundaries [21–31]. The open-source quantum mechanical simulation package "DFTB+" [17,32] is used in our simulations.

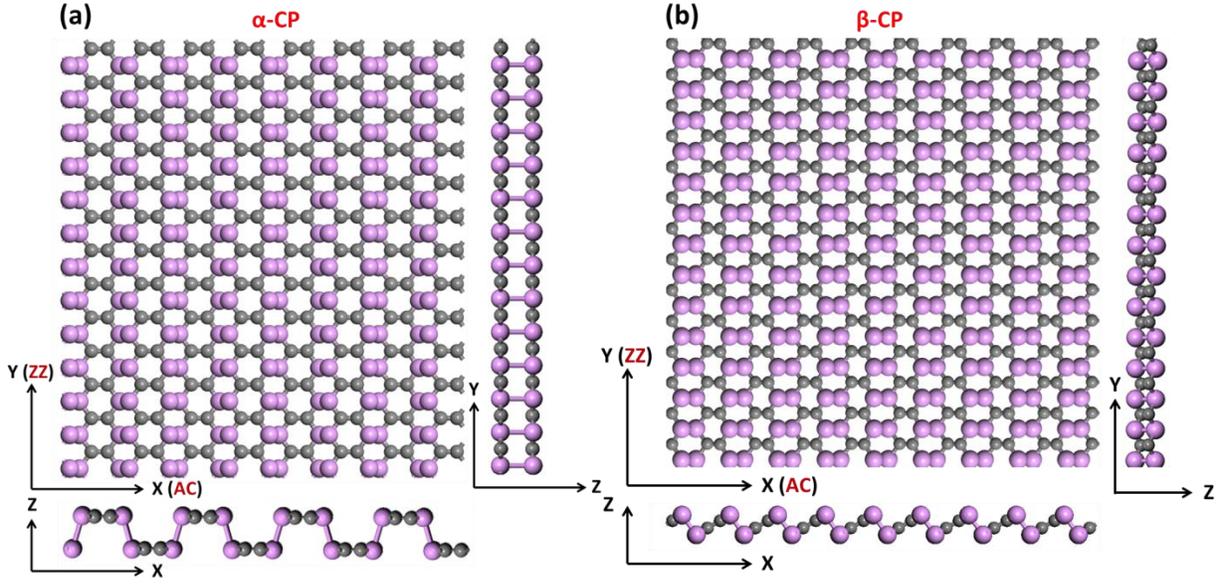

**Figure 1:** Two types of carbon phosphide (CP) monolayers, α-CP (a) and β-CP (b), with the arm-chair (AC) direction oriented along the X-axis and the zig-zag (ZZ) direction oriented along the Y-axis. The top (XY-pane), the side (XZ-plane) and the front (YZ-plane) are shown. Carbon atoms are grey and phosphorous atoms are purple.

First, we optimized the unit cells of two allotropes of carbon phosphide, which were previously obtained by the DFT method [1]. We found that the overall deviation in the bond length and the bond angles calculated by DFTB from those obtained by the DFT does not exceed ~15%, in accordance with the earlier DFTB simulations [21–31]. The monolayers of α-CP (see Figure 1(a)) and β-CP (see Figure 1(b)) were constructed, and then carbon, phosphorous, and mixed monovacancies and divacancies were created by taking away a single atom or a couple of neighboring atoms (see Figure 2). The initial atomic configurations containing the hosted vacancies were optimized by minimizing the total energy of the systems. The relaxed atomic configurations are shown in Figure 3. The size of the CP monolayers is specified by the number of unit cells $n_{AC}$ along AC direction (oriented along the X–axis) and the number of unit cells $n_{ZZ}$ along ZZ direction (oriented along the Y–axis). For the monolayer of α-CP, we set $n_{AC} = 8$ and $n_{ZZ} = 13$ with the corresponding length $L_{AC} = 34.35$ Å along AC direction, and $L_{ZZ} = 38.64$ Å along ZZ direction (see Figure 1(a)). For the monolayer of β-CP, we set $n_{AC} = 9$ and $n_{ZZ} = 14$ with the resultant length $L_{AC} = 39.02$ Å along AC direction, and $L_{ZZ} = 41.99$ Å along ZZ direction (see Figure 1(b)). The total number of atoms in the α-CP monolayer is *N*= 416 atoms and that in the β-CP monolayer is *N*=414 atoms.

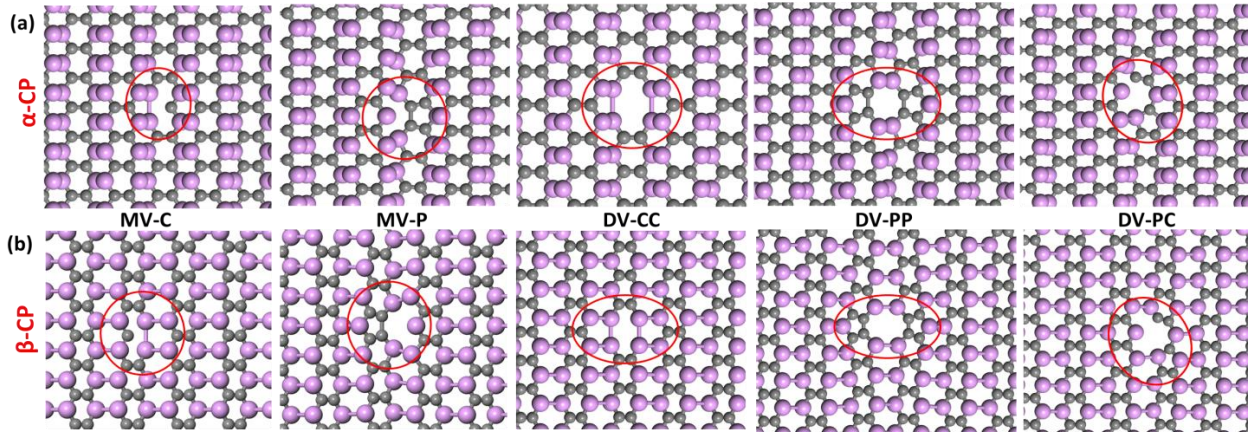

Figure 2: Initial configuration of the constructed single and double vacancies in α-CP (a) and β-CP (b) monolayers. From the left to the right: carbon monovacancy (MV-C), phosphorous monovacancy (MV-P), carbon divacancy (DV-CC), phosphorous divacancy (MV-PP) and mixed phosphorus-carbon divacancy (DV-PC). Carbon atoms are grey and phosphorous atoms are purple. The ellipses around the vacancies are to guide the eye.

We studied the mechanical response and the failure of pristine and defective allotropes of carbon phosphide subjected to uniform uniaxial tensile strain applied along AC and ZZ directions. Periodic boundary conditions were applied along the direction orthogonal to the loading direction. In order to apply periodic boundary conditions along the Z-axis, a vacuum slab with the thickness of Δ=40 Å was added to eliminate the self-interaction of the monolayers (due to the periodic boundary conditions applied along the Z-axis). The k-point set for the Brillouin-zone integration was chosen using the Monkhorst-Pack technique [33]. The Monkhorst-Pack grid [33] with a 2x2x2 sampling set was applied for Brillouin-zone integration. In accordance with the previous DFTB studies of graphene [18–20] and phosphorene [21–31], the s- and p-orbitals were specified for each carbon atom (C-atom) and phosphorous atom (P-atom). The Slater-Koster files [9] for C- and P-atoms were taken from the 'MATSCI' set [34,35].

We applied quasi-statically uniaxial tensile strain to the carbon phosphide monolayers along the selected primary direction at zero temperature. The tensile strain was gradually incremented by $\delta\varepsilon$=0.01. Consequently, the geometry of the defect-free and defective monolayers was optimized by minimizing the total energy at a given tensile strain by using the conjugate gradient method [36]. The SCC calculations were implemented at every step of the energy minimization with the DFTB method. The nominal stress was obtained as explained in [37] by calculating the first derivative of the energy density (the total energy divided by the volume of the CP monolayer). The volume of the CP monolayer was calculated as a product of its area ($A = L_{AC} \cdot L_{ZZ}$) and its effective thickness, $w$. The effective thickness for the monolayer was taken as the inter-layer distance in α-CP and β-CP 3D crystals. Using the unit cells of the two CP allotropes, we applied periodic boundary conditions in all three directions and obtained the 3D lattice structure carbon phosphide crystals with the DFTB+ method. The obtained effective thickness for α-CP monolayer is $w_\alpha = 5.75$ Å and for β-CP monolayer, it is $w_\beta$ = 6.35 Å.

# 3. Results and Discussion

## 3.1 Defect structure and energetics

*Structure of vacancies*

First we optimized the structure of constructed CP monolayers with the hosted vacancies by using energy minimization. The obtained atomistic configurations around the introduced vacancies are shown in Figure 3(a) for α-CP and in Figure 3(b) for β-CP monolayers, respectively. The atomistic configurations around a carbon monovacancy and a divacancy in α-CP are rather similar to those in graphene and phosphorene [38,39]. The atomistic configuration of the carbon monovacancy (MV-C) is formed by two adjacent pentagon-nonagon (5|9) atomic rings containing P- and C-atoms. An extra phosphorous-phosphorous (PP) bond is found between a pair of neighboring P-atoms in the nonagon ring (near the highlighted C-atom of the MV-C vacancy in Figure 3(a)). The carbon divacancy (DV-CC) has the most energetically favorable atomistic configuration consisting of three adjacent pentagon-octagon-pentagon (5|8|5) atomic rings (see the DV-CC divacancy in Figure 3(a)). In contrast to the monovacancy, where the highlighted C-atom has a single dangling bond, there are no dangling bonds around the vacant site of the DV-CC divacancy.

The spatial energy distributions around the MV-C and the DV-CC vacancies are shown in Figure 4(c, d). The atoms with the highest energy are the P-atoms located at the vacancy cores. This high energy is associated with the strain energy of the deformed PP bonds found around the vacancy sites: the length of the PP bond shared by the pentagon-octagon rings in the MV-C is shortened by $\frac{\Delta l}{l_0}$ =-1.5% compared to the PP bond length ($l_0$) in the bulk of pristine α-CP, while in the DV-CC, the PP bonds shared by the pentagon-octagon-pentagon rings in MV-C are compressed by $\frac{\Delta l}{l_0}$ = -6.8%.

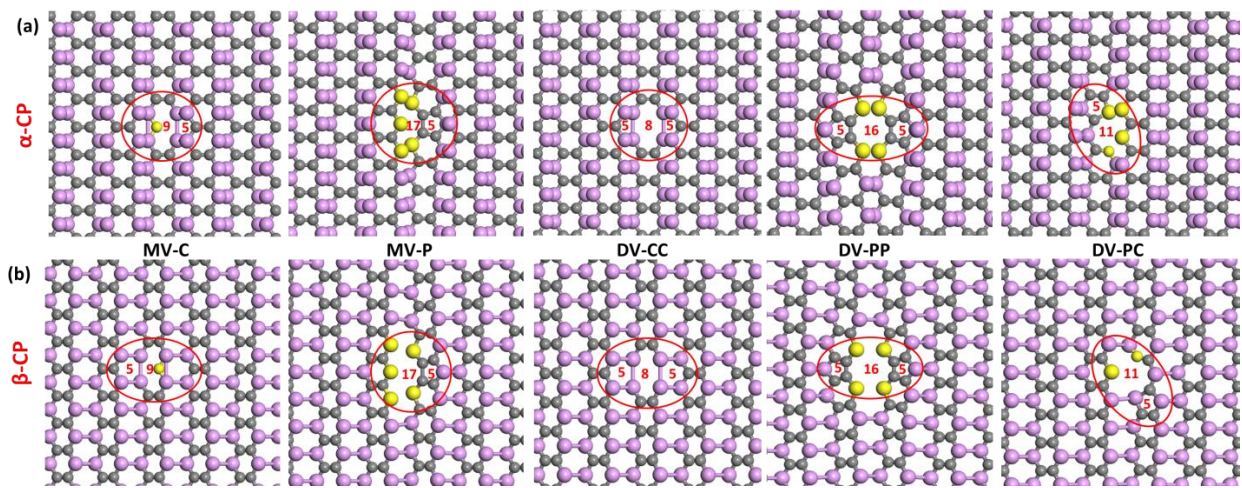

Figure 3: Optimized atomistic configurations of single and double vacancies in α-CP (a) and β-CP (b) monolayers obtained by energy minimization. From the left to the right: carbon monovacancy (MV-C), phosphorous monovacancy (MV-P), carbon divacancy (DV-CC), phosphorous divacancy (MV-PP) and phosphorus-carbon divacancy (DV-PC). Carbon atoms are grey and phosphorous atoms are purple. The atoms with dangling bonds are highlighted. The ellipses around the vacancies are to guide the eye. The overlaid numbers indicate the number of atoms in the adjacent atomic rings forming vacancies.

The atomistic configurations around a MV-P monovacancy and a DV-PP divacancy are shown in Figure 3(a). In contrast to C-atoms, the elimination of P-atoms disturbs significantly the lattice structure around a vacant site. The removal of a single P-atom leads to the formation of an extended MV-P monovacancy with a (5|17) atomistic structure, while the deletion of a couple of adjacent P-atoms results in an expanded (5|16|5) DV-PP divacancy (see Figure 3(a)). The formation of phosphorous vacancies is accompanied by bond breaking between the nearby P-atoms (see the P-atoms highlighted in Figure 3(a)). The elimination of a P-atom leads to a subtle bond misbalance near the vacant site: when a single P-atom is removed, the adjacent C-atoms (located in the same plane as the removed P-atom) move closer to each other and form a strong covalent CC bond. This pair of the C-atoms displaces alongside their adjacent P-atoms (to which they are linked via the CP bonds). Consequently, the PP bonds between the adjacent P-atoms rupture since PP bonds are considerably weaker than CP and CC bonds. As a result, the five P-atoms with dangling bonds appear in the MV-P monovacancy, and in the same way, the four P-atoms with dangling bonds emerge in the DV-PP divacancy (see the highlighted P-atoms in Figure 3(a)). The presence of the P-atoms with dangling ponds is manifested in the high formation energy of phosphorous vacancies, which has a strong detrimental effect on the mechanical properties of CP monolayers.

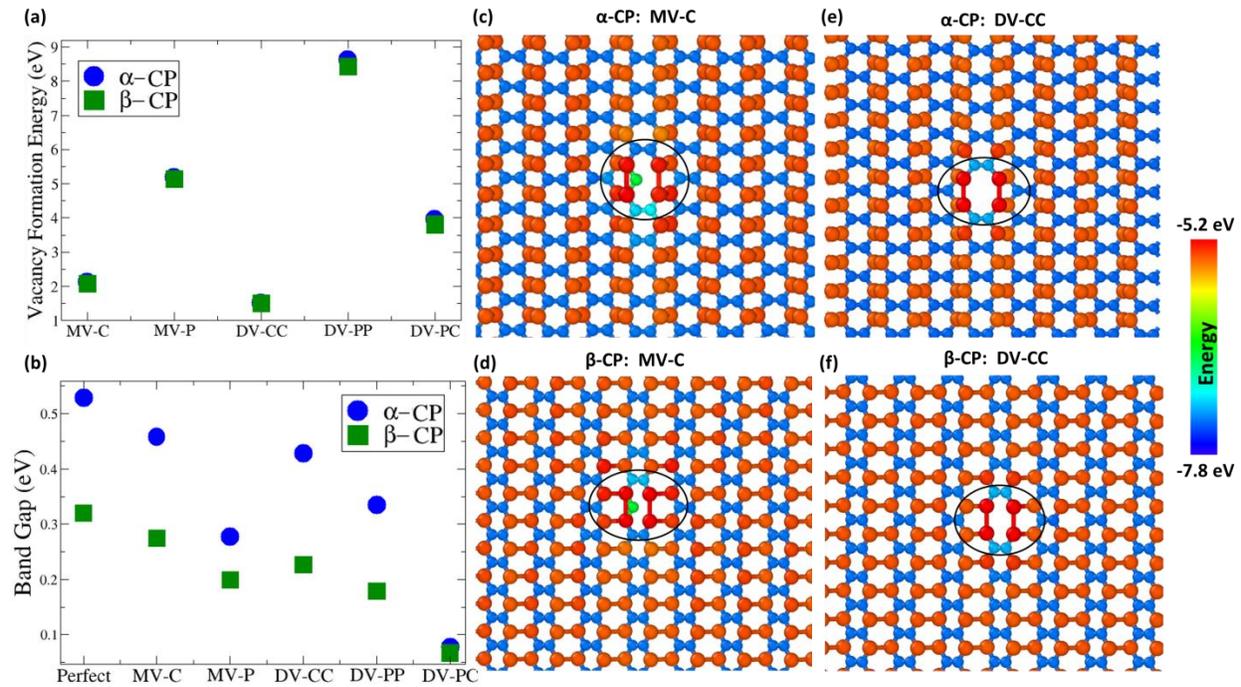

Figure 4: (a) Formation energies of vacancies in α-CP (blue circles) and β-CP (green squares) monolayers. (b) Band gaps of α-CP (blue circles) and β-CP (green squares) monolayers. (d-e) Spatial energy distributions around a monovacancy MV-C in α-CP (d) and β-CP(c) monolayer, and a divacancy DV-CC in α-CP (e) and β-CP (f) monolayer. The atomistic images are taken at the reference configuration at zero strain. Energy per atom is specified by color. The energy variation is specified by the color bar on the right.

Besides carbon and phosphorous divacancies, a new type of mixed phosphorous-carbon divacancies (DV-PC) can be introduced by removal of a pair of adjacent P- and C-atoms located in the same (upper or lower) atomic plane of CP monolayer. When the CP pair is removed, the remaining three of their four nearest neighboring atoms (one P-atom and two C-atoms) are located in the same atomic plane, while the fourth neighboring P-atom lies in a different plane. Each of the four atoms has a dangling bond. After energy relaxation, a new CP bond can be formed between two of the three atoms located in the same plane since the initial distance between them (before the energy relaxation) is relatively short (d=2.8Å). However, the other pair of P- and C-atoms located in different atomic planes remain unbonded since the initial distance between them is excessively large (d=3.6 Å, see the highlighted C- and P-atoms around the DV-PC vacancy in Figure 3(a)). In addition, the formation of the new CP bond causes the displacement of the adjacent P-atoms and subsequent rupture of the weak PP bonds between them. Hence, an additional pair of P-atoms with dangling bonds appears in the divacancy core (see the highlighted pair of adjacent P-atoms around the DV-PC vacancy in Figure 3(a)). The resultant mixed DV-PC divacancy has an extended (5|11) structure formed by two adjacent 5- and 11-membered atomic rings as shown in Figure 3(a). Thus, removal of a single P-atom or a pair of adjacent P-atoms results in appearance of the extra P-atoms with dangling bonds around phosphorous vacancies, and additional P- and C-atoms with dangling bonds around mixed divacancies.

The removal of P- and C-atoms from the β-CP monolayer results in the almost identical structure of pure and mixed mono- and divacancies (see Figure 3(b)). The spatial energy distributions around carbon vacancies are shown in Figure 4(e, f). In a similar way, the atoms with the higher than average atomic energy are located in the vacancy core. The similarity in the vacancy structures and the spatial energy distributions around the defects in the two carbon phosphide allotropes is remarkable.

### *Vacancy formation energy*

We calculated the vacancy formation energy as $E_{vac} = E_{tot} - \frac{N}{N_p} E_{p,tot}$, where $E_{tot}$ is the potential energy of a defective CP monolayer containing *N* atoms, while $E_{p,tot}$ is the potential energy of the equivalent pristine (defect-free) monolayer with $N_p$ atoms ($N = N_p - n$, where *n* is the number of removed atoms: *n*=1 for monovacancy and *n*=2 for divacancy).

The calculated vacancy formation energies for both α-CP and β-CP monolayers are represented in Figure 4(a) (see also Table 1). The vacancy formation energies in α-CP and β-CP monolayers are comparable. As can be seen in Figure 4(a), a carbon divacancy (DV-CC) has the lowest formation energy of $E_{vac}\sim1.5$ eV, while the formation energy of a carbon monovacancy (MV-C) is somewhat higher, with $E_{vac}\sim2.1$ eV. The phosphorus divacancy has the highest formation energy of $E_{vac}\sim8.6$ eV. For comparison, the vacancy formation energy in phosphorene is $E_{vac}\sim1.6$ eV for a monovacancy and $E_{vac}\sim1.4$ eV for a (5|8|5) divacancy [40], while in graphene, $E_{vac}\sim7.6$eV for a monovacancy and $E_{vac}\sim9.0$ eV for a (5|8|5) divacancy [41,42], respectively.

**Table 1: Vacancy formation energies and band gaps in pristine and defective α and β CP monolayers.**

| Defect | Vacancy formation energy (eV) | | Band gap (eV) | |
|---|---|---|---|---|
| | α-CP | β-CP | α-CP | β-CP |
| Perfect | - | - | 0.53 | 0.32 |
| MV-C | 2.13 | 2.08 | 0.46 | 0.28 |
| MV-P | 5.18 | 5.14 | 0.28 | 0.20 |
| DV-CC | 1.52 | 1.50 | 0.43 | 0.23 |
| DV-PP | 8.64 | 8. 43 | 0.34 | 0.19 |
| DV-PC | 3.96 | 3.81 | 0.09 | 0.07 |

In addition, the band gaps of pristine and defective CP monolayers were calculated (see Figure 4(b) and Table 1). As shown in Figure 4(b), the band gap of pristine α-CP monolayer ($E_{bg}$= 0.63 eV) is larger than that of β-CP ($E_{bg}$= 0.32 eV). The band gap for pristine α-CP and β-CP monolayers calculated with the DFTB method are comparable to those obtained from DFT: Singh et al. [2] obtained $E_{bg}$= 0.63 eV for α-CP and $E_{bg}$= 0.36 eV for β-CP by using GGA-PBE functional, while Wang et al. [1] obtained $E_{bg}$= 1.26 eV for α-CP and $E_{bg}$= 0.87 eV for β-CP by using more accurate HSE06 density functional. The band gaps in defective α and β carbon CP monolayers are reduced by the hosted vacancies. The major reduction is due to phosphorous and mixed vacancies, while the effect of carbon vacancies is relatively weaker.

## 3.2 Young's modulus and Poisson's ratio

### Young's Modulus

Next, we investigated the effect of the hosted vacancies on the mechanical properties (Young's modulus and Poisson's ratio) of carbon phosphide monolayers. The Young's modulus (Y) of CP monolayers was calculated as the second derivative of the total energy with respect to applied strain (in the reference configuration at zero strain, ε=0), according to:

$$Y = \frac{1}{V_0}\left(\frac{\partial^2 E_{PNT}}{\partial \varepsilon^2}\right)_{\varepsilon=0}$$

where $V_0$ is the volume of the CP monolayer measured at zero strain. The second derivative was obtained by subjecting the α- and β-CP monolayers to infinitesimally small compressive and tensile strain (-3% ≤ ε ≤ 3%), with following optimization of their geometries at a given strain. Then the calculated total energy as a function of uniaxial strain was fitted to a second order polynomial. The polynomial was used to obtain the second derivative.

The Young's modulus (see Figure 5(a-d)) was calculated along AC and ZZ directions for pristine and defective CP monolayers (see also Table 3 for α-CP and Table 4 for β-CP, correspondingly). We found a strong anisotropy in the mechanical properties of pristine carbon phosphide. The Young's modulus calculated along the ZZ direction in α-CP ($Y_{ZZ}$=343.04 GPa) is by one order of magnitude larger than that along the AC direction ($Y_{AC}$=35.5 GPa). Likewise, the Young's modulus along the ZZ direction in β-CP ($Y_{ZZ}$=288.85 GPa) is also larger than that along the AC direction ($Y_{AC}$=81.25 GPa). Similar to phosphorene, the strong anisotropy is due to the puckered structure of carbon phosphide, which adapts to the applied tensile strains distinctively differently along the two directions. The pronounced anisotropy is also related to the reinforcing character of the strong covalent CC bonds (and to some extent, relatively weaker CP bonds). We also note that α-CP monolayer is stiffer along the ZZ direction than the β-CP one, while the β-CP is stiffer along the AC direction than α-CP one.

The obtained Young's moduli along the principal directions are similar to those calculated at room temperature using DFT method by Wang et al. [1]. More specifically, the in-plane stiffness obtained from DFT calculations along the AC direction: $C_{AC}$=18.75 N/m for α-CP and $C_{AC}$=45.56 N/m for β-CP [1] are close to the values obtained at 0 temperature by DFTB: $C_{AC}$=20.3 N/m for α-CP and $C_{AC}$=51.6 N/m for β-CP. The in-plane stiffness evaluated by DFT in the ZZ direction: $C_{ZZ}$=171.47 N/m for α-CP and $C_{ZZ}$=158.27 N/m for β-CP [1] are also comparable to the values obtained at T=0K by DFTB: $C_{ZZ}$=196.2 N/m for α-CP and $C_{ZZ}$=173.42 N/m for β-CP.

The Young's moduli of α-CP and β-CP monolayers along the ZZ direction are significantly larger than that of phosphorene ($Y_{ZZ}$=166.2GPa). Along the AC direction, the Young's modulus of β-CP is twice as large as that of phosphorene ($Y_{AC}$=44.6GPa) [43], while the Young's modulus of α-CP is comparable to that of phosphorene. In comparison, the Young's modulus of graphene is much higher ($Y_{ZZ} = Y_{AC} = 1000$ GPa) [44].

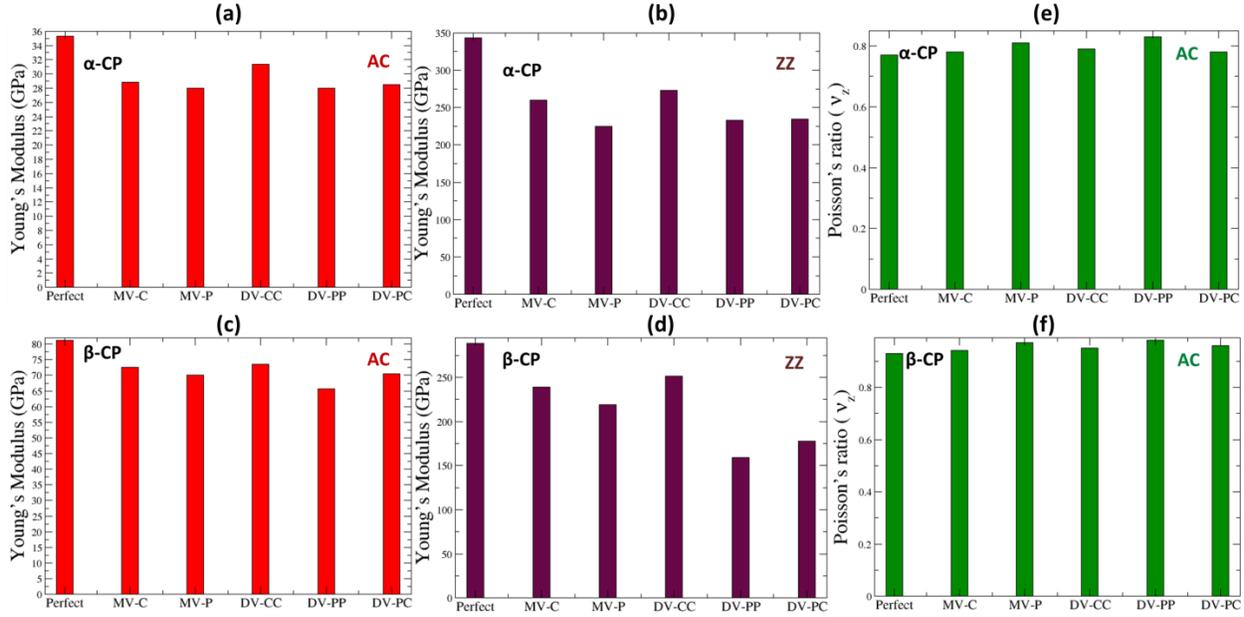

Figure 5: Mechanical properties of pristine and defective carbon phosphide: the Young's modulus calculated along the AC direction for α-CP (a) and β-CP (c) monolayer and along the ZZ direction for α-CP (b) and β-CP (d) monolayer. The Poisson's ratio $v_z$ calculated for α-CP (e) and β-CP (f) monolayer stretched along the AC direction.

The Young's modulus of α and β carbon phosphide monolayers can be noticeably reduced by the embedded vacancies (see in Figure 5(a-d)). Along the AC direction, the strongest detrimental effect in α-CP is due to phosphorous vacancies: a DV-PP divacancy and a MV-C monovacancy both reduce the Young's modulus by $\frac{\Delta Y_{AC}}{Y_{AC}^p} \approx$ -21% ( $\Delta Y_{AC} = Y_{AC} - Y_{AC}^p$, where $Y_{AC}^p$ is the Young's modulus of pristine α-CP monolayer along the AC direction). Carbon divacancies have a relatively weaker impact on the Young's modulus: a DV-CC divacancy reduces the Young's modulus only by $\frac{\Delta Y_{AC}}{Y_{AC}^p} \approx$ -11%. The effects of a MV-C monovacancy ($\frac{\Delta Y_{AC}}{Y_{AC}^p} \approx$ -18%) and a mixed DV-PC divacancy ($\frac{\Delta Y_{AC}}{Y_{AC}^p} \approx$ -19%) are comparable to those of phosphorous vacancies.

Similarly, the detrimental effect of phosphorous vacancies on the Young's modulus of α-CP monolayer is the most pronounced along the ZZ direction: a MV-P monovacancy reduces $Y_{ZZ}$ by $\frac{\Delta Y_{ZZ}}{Y_{ZZ}^p}$ -34% while a DV-PP divacancy and a mixed DV-PC divacancy decrease $Y_{ZZ}$ by $\frac{\Delta Y_{ZZ}}{Y_{ZZ}^p} \approx$ -13%. The effect of carbon vacancies on the Young's modulus is comparatively weaker: a MV-C monovacancy reduces $Y_{ZZ}$ by $\frac{\Delta Y_{ZZ}}{Y_{ZZ}^p} \approx$ -24%, while a DV-PP divacancy decreases $Y_{ZZ}$ by $\frac{\Delta Y_{ZZ}}{Y_{ZZ}^p} \approx$ -20%.

In the same way, the detrimental effect of phosphorous vacancies on the Young's modulus of β-CP monolayer is the strongest one: a DV-PP monovacancy reduces it by $\frac{\Delta Y_{AC}}{Y_{AC}^p} \approx$ -19% along the AC direction, and by $\frac{\Delta Y_{ZZ}}{Y_{ZZ}^p} \approx$ -45% along the ZZ direction. A MV-P monovacancy reduces the Young's modulus by $\frac{\Delta Y_{AC}}{Y_{AC}^p} \approx$ -14% and by $\frac{\Delta Y_{ZZ}}{Y_{ZZ}^p} \approx$ -24% along the AC and ZZ directions, respectively. A mixed DV-PC divacancy reduces it by $\frac{\Delta Y_{AC}}{Y_{AC}^p} \approx$ -13% and $\frac{\Delta Y_{ZZ}}{Y_{ZZ}^p} \approx$ -38% along the AC and ZZ directions, respectively.

Similarly to α-CP monolayer, the impact of carbon vacancies on Young's modulus of β-CP monolayer is relatively moderate: a MV-C monovacancy reduces Young's modulus only by $\frac{\Delta Y_{AC}}{Y_{AC}^p} \approx$ -11% along the AC direction, and by $\frac{\Delta Y_{ZZ}}{Y_{ZZ}^p} \approx$ -17% along the ZZ direction; whereas a DV-CC divacancy decreases it by $\frac{\Delta Y_{AC}}{Y_{AC}^p} \approx$ -9% and $\frac{\Delta Y_{ZZ}}{Y_{ZZ}^p} \approx$ -13% along the AC and ZZ directions, respectively. The Young's modulus of carbon phosphide is noticeably less affected by vacancies along the AC direction than along the ZZ direction due to its specific puckered structure. This is because for the AC direction, the relatively weak PP bonds between P-atoms (oriented along the AC direction) and the corresponding bond angles in the puckers can effectively rearrange around a vacancy to minimize the negative effect on the Young's modulus.

The sensitivity of the Young's modulus to the hosted vacancies is comparable for α-CP and β-CP allotropes. Both allotropes show almost equivalent reduction in the Young's modulus caused by the accommodated vacancies along the AC direction ($\frac{\Delta Y_{AC}}{Y_{AC}^p} \approx$ -18%). In contrast, along the ZZ direction the Young's modulus of the β-CP monolayer is more affected by phosphorous divacancies ($\frac{\Delta Y_{ZZ}}{Y_{ZZ}^p} \approx$ -45%) than $Y_{ZZ}$ of the α-CP monolayer ($\frac{\Delta Y_{ZZ}}{Y_{ZZ}^p} \approx$ -34%). Graphene and phosphorene are also softened by vacancies [45–47]: monovacancies reduce the Young's modulus of phosphorene along the AC direction by $\frac{\Delta Y_{AC}}{Y_{AC}^p} \approx$ -10% [46] and along the ZZ direction by $\frac{\Delta Y_{ZZ}}{Y_{ZZ}^p} \approx$ -12% [47], whereas the Young's modulus of graphene is reduced by $\frac{\Delta Y}{Y^p} \approx$ -10% at low vacancy concentration [45]. The impact of phosphorene vacancies (and especially divacancies) on the Young's modulus is substantially larger in carbon phosphide than in phosphorene or graphene.

### *Poisson's ratio*

We calculated the Poisson's ratio that relates the resulting transverse strain to applied uniaxial tensile strain. Under the tensile strain applied along the X-axis, the CP monolayers contract in two mutually orthogonal transverse directions: along the Y-axis (in-plane) and the Z-axis (out-of-plane). The Poisson's ratio along the Y-direction was calculated according to:

$$\frac{\Delta L_y}{L_y^0} = -\nu_y^\gamma \frac{\Delta L_x}{L_x^0},$$

where $L_x$ ($L_y$) is the length of the computational box along the X-axis (Y-axis) at a given tensile strain, while $L_x^0$ ($L_y^0$) is the reference box length along the X-axis (Y-axis) at zero strain, and $\gamma$ indicates either the AC or the ZZ direction of the applied tensile strain.

In a similar way, the Poisson's ratio $\nu_z^\gamma$ along the Z-direction was obtained as:

$$\frac{\Delta w}{w^0} = -\nu_z \frac{\Delta L_x}{L_x^0},$$

where $w$ is the thickness of the monolayer at a given tensile strain, while $w^0$ is the reference thickness at zero strain, and $\gamma$ indicates either the AC or the ZZ direction of the applied tensile strain

The calculated Poisson's ratios $v_y^Y$ and $v_z^Y$ for pristine and defective CP monolayers are reported in Table 2. The strong anisotropy in the mechanical properties is equally revealed in the Poisson's ratios. The Poisson's ratios obtained under stretching along the AC and the ZZ directions are fairly different, which reflects the characteristic anisotropy of the puckered structure of α and β CP allotropes. In both α-CP and β-CP monolayers, $v_y^{AC}$ is significantly smaller than $v_y^{ZZ}$ at least by an order of magnitude, while $v_z^{AC}$ and $v_z^{ZZ}$ are sufficiently dissimilar (see Table 2). The calculated in-plane and the out-of-plane Poisson's ratios are also different by one order of magnitude. The highly anisotropic Poisson's ratio of CP monolayers is similar to that of phosphorene. In phosphorene, the Poisson's ratio is $v_y^{AC}$=0.05 and $v_y^{ZZ}$=0.93 [48] along the AC and the ZZ directions, respectively. This strong anisotropy in the Poisson's ratios along the primary directions is determined by the puckered structures of phosphorene and CP monolayers.

Introduction of vacancies only moderately affects the Poisson's ratios of CP monolayers (see Figure 5(e, f)). The overall increase in the magnitude of the Poisson's ratio is comparable to that previously found in phosphorene nanoribbons [49] and nanotubes [31]. The softening of CP monolayers due to the accommodated vacancies leads to enhanced compressibility in the in-plane and the out-of-plane transverse directions, and therefore to an increase in the Poisson's ratios (see Table 2). The Poisson's ratios along the AC direction $v_y^{AC}$ and $v_z^{AC}$ of α-CP monolayer are mostly affected by phosphorous divacancies: their values increase by ~10% and ~8%, correspondingly, with respect to a defect-free monolayer. Equally, the Poisson's ratios $v_y^{ZZ}$ and $v_z^{ZZ}$ (along the ZZ direction) increase by ~8% and ~10%, respectively, due to phosphorous divacancies. In contrast, the Poisson's ratios are least affected by the carbon vacancies: the overall increase in the magnitudes is less than ~1%. To the same extent, the Poisson's ratios along the AC direction $v_y^{AC}$ and $v_z^{AC}$ of β-CP monolayer are also mainly affected by the hosted phosphorous divacancies: their magnitude increases by ~10% and ~5%, respectively. The Poisson's ratios along the ZZ direction $v_y^{ZZ}$ and $v_z^{ZZ}$ are enlarged by ~11% and ~8%, respectively, due to phosphorous divacancies. Like in α-CP monolayer, the Poisson's ratios in β-CP monolayer are least affected by carbon vacancies: the overall increase in their values is less than ~2% (see Table 2).

Table 2: Poisson's ratios for defect-free and defective α-CP and β-CP monolayers calculated along the transverse directions (Y- and the Z-axes) when the axial direction (X-axis) oriented along the AC or ZZ directions.

| Poisson's ratio | α-CP (AC) | | α-CP (ZZ) | | β-CP (AC) | | β-CP (ZZ) | |
|---|---|---|---|---|---|---|---|---|
| | $v_y$ | $v_z$ | $v_y$ | $v_z$ | $v_y$ | $v_z$ | $v_y$ | $v_z$ |
| Perfect | 0.020 | 0.77 | 0.46 | 0.42 | 0.058 | 0.93 | 0.35 | 0.52 |
| MV-C | 0.020 | 0.78 | 0.47 | 0.43 | 0.059 | 0.94 | 0.36 | 0.53 |
| MV-P | 0.022 | 0.81 | 0.47 | 0.43 | 0.062 | 0.97 | 0.38 | 0.55 |
| DV-CC | 0.021 | 0.79 | 0.48 | 0.45 | 0.060 | 0.95 | 0.37 | 0.54 |
| DV-PP | 0.022 | 0.83 | 0.49 | 0.46 | 0.064 | 0.98 | 0.39 | 0.56 |
| DV-PC | 0.021 | 0.78 | 0.47 | 0.44 | 0.061 | 0.96 | 0.36 | 0.54 |

## 3.3 Tensile deformation and failure

### Deformation and failure of pristine CP monolayers

Finally we studied the tensile deformation and failure of pristine and defective CP monolayers under uniaxial tensile straining.

### *Uniaxial deformation*

Pristine monolayers of α and β allotropes were uniaxially stretched along the AC and the ZZ directions up to the failure strain, at which the sample lost its structural stability and failed. The strain energies as a function of applied tensile strain for α-CP and β-CP allotropes are shown in Figure 8(a, b), respectively and the stress-strain curves are shown in Figure 8(c, d), respectively. The remarkable anisotropy in the mechanical properties of CP monolayers is evident in Figure 8(c, d): the stress-strain curve is much steeper along the ZZ direction than along the AC direction, and the failure stress is significantly larger along the ZZ direction.

At a small strain ($\varepsilon \lesssim 0.3$), the tensile deformation along the AC direction proceeds via flattening of the puckered structure of α- and β-CP monolayers. The flattening occurs via a hinge-like rotation of the PP bonds oriented along the AC direction, and it is accompanied by the moderate stretching of the PP bonds. At larger strain ($\varepsilon > 0.3$), the tensile deformation is mostly due to stretching of the PP bonds (and to a less extent of the CC and CP bonds). The investigated monolayers can be substantially strained along the AC direction thanks to the flattening of the puckered structure via the bond rotation. In contrast, the tensile deformation along the ZZ direction is accomplished by direct stretching of the CP bonds, which are aligned at an acute angle to the direction of the applied tensile strain. These bonds eventually ruptured, causing the structure failure.

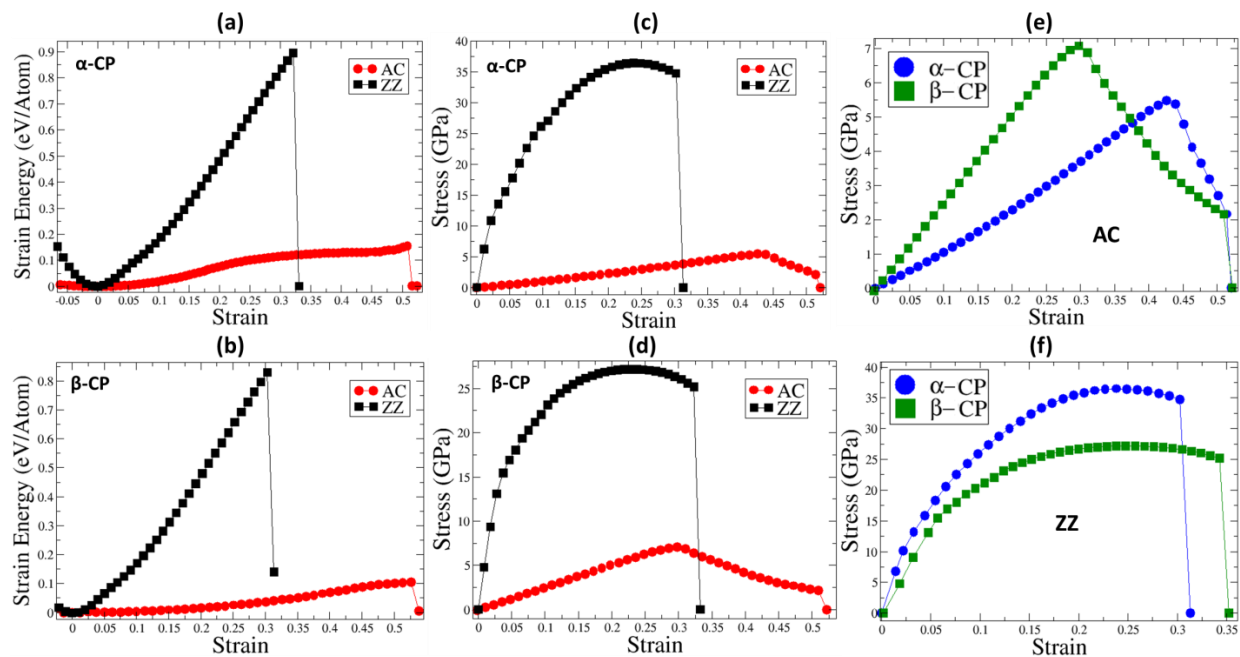

Figure 6: Strain energy as a function of uniaxial tensile strain applied along the AC (red circles) and the ZZ (black squares) directions for α-CP (a) and β-CP (b) monolayers. Stress-strain curves calculated for α-CP (c) and β-CP (d) monolayers stretched along the AC (red circles) and the ZZ (black squares) directions. Comparison of the stress-strain curves for α-CP (blue circles) and β-CP (green squares) monolayers calculated along the AC (e) and the ZZ (f) directions.

## Failure Mechanism

The gradual tensile straining of CP monolayers inevitably leads to the failure as soon as a critical tensile strain is reached. However, the failure mechanism of a pristine CP monolayer stretched along the AC and ZZ directions are rather different as shown in Figure 7. This is another manifestation of the characteristic anisotropy in the mechanical properties related to the puckered structure of CP monolayers.

The failure of pristine α-CP monolayer stretched along the AC direction is illustrated in Figure 7(a). At a large strain, the weak PP bonds in α-CP are the most stretched bonds, which fracture first at a critical strain; while the stronger CP and CC bonds remain intact (see Figure 7(a)). As a result, the structural stability is lost at the critical strain and the sample disintegrates, forming a set of disconnected fragments. The critical tensile strain at which α-CP fails along the AC direction is $\varepsilon_{AC}^p$=0.54. The failure mechanism of β-CP is almost identical, at the fracture strain of $\varepsilon_{AC}^p$=0.52. These failure strains of CP allotropes are fairly close to that of phosphorene $\varepsilon_{AC}^p$=0.5 [43]. The fracture stresses along the AC direction are $\sigma_{AC}^p$=5.49GPa for α-CP and $\sigma_{AC}^p$=7.15GPa for β-CP, respectively, which are also comparable with that of phosphorene $\sigma_{AC}^p$=8.1GPa [50].

When pristine α-CP monolayer is stretched along the ZZ direction, it fails in a different way (see Figure 7(b)). The tensile deformation is accomplished by stretching of the CP bonds oriented at an acute angle to the ZZ direction (the PP and CC bonds are oriented normally to that direction). At a critical strain of $\varepsilon_{ZZ}^p$=0.32, the CP bonds (and also a small fraction of the PP bonds) rupture throughout the entire sample. As a result, the α-CP monolayer falls apart via fragmentation, forming a set of short strings consisting of the C- and P-atoms (see Figure 7(b)). The same failure mechanism was also observed for pristine β-CP monolayer uniaxially strained along the ZZ direction. Since CP bonds are evidently stronger than PP bonds, the fracture strain along the ZZ direction for α-CP $\varepsilon_{ZZ}^p$=0.32 and β-CP $\varepsilon_{ZZ}^p$=0.34 allotropes exceeds the failure strain of pristine phosphorene $\varepsilon_{ZZ}^p$=0.24 [43]. Similarly, the fracture stresses along the ZZ direction for α-CP monolayer $\sigma_{ZZ}^p$=36.5GPa and β-CP monolayer $\sigma_{ZZ}^p$=27.5GPa are larger than that of phosphorene $\sigma_{ZZ}^p$=18GPa [50].

It is worth noting again the strong anisotropy in the mechanical properties of both CP allotropes: the failure strain along the AC direction is almost twice of that along the ZZ direction, while the failure strain along the ZZ direction exceeds by an order of magnitude of that along the AC direction (see Table 3 for α-CP and Table 4 for β-CP, respectively)

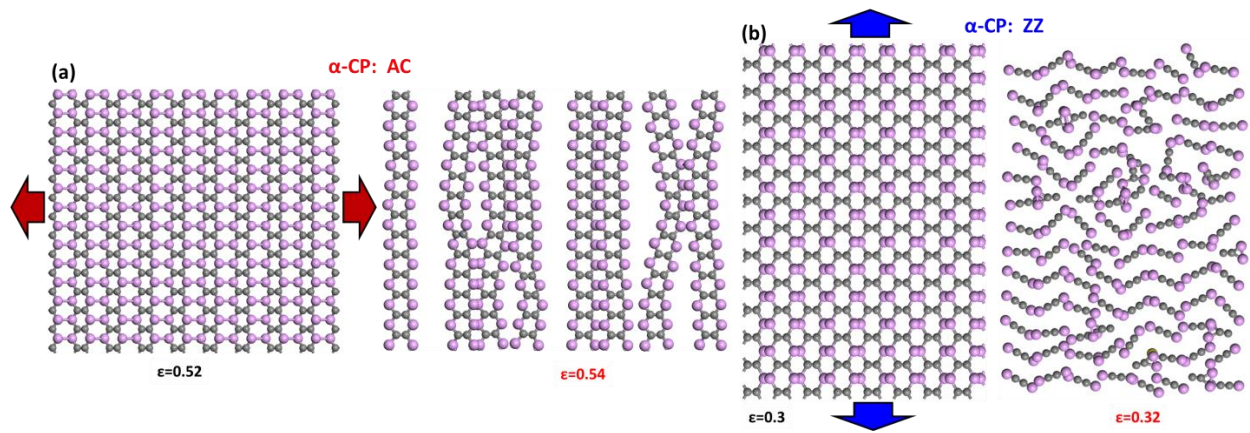

**Figure 7:** (a) Failure of pristine α-CP monolayer stretched along the AC direction. Two successive atomistic configurations are shown before (ε=0.52) and after ($\varepsilon_f$=0.54) failure, respectively. (a) Failure of pristine α-CP monolayer under uniaxial tensile strain applied along the ZZ direction. Two consecutive atomistic configurations are taken before (ε=0.3) and after ($\varepsilon_f$=0.32) failure, correspondingly. Arrows indicate the direction of applied tensile strain. Carbon atoms are grey and phosphorous atoms are purple.

## Deformation and failure of defective α-CP monolayers

We next investigated the deformation and failure mechanism of defective CP containing single and double vacancies. Like before, a defective monolayer was uniaxially stretched along the AC and ZZ directions. In what follows, for clarity sake, the detailed analysis of the deformation and the failure mechanism is presented only for a few selected types of single and double vacancies.

The obtained stress-strain curves for a defective α-CP monolayer stretched along the AC and ZZ directions are shown in Figure 8(a, b), respectively. The stress-strain curve of pristine α-CP monolayer is plotted for comparison. As can be seen from Figure 8(a, b), the failure strain and the failure stress of the defective sample are strongly affected by the hosted vacancies.

### *Failure along the AC direction*

The extent of the effect depends on the type of vacancy: the effect of phosphorous vacancies is the strongest one. Due to the structural anisotropy, the deformation and failure mechanism also depends on the stretching direction. In Figure 10(a), we present the deformation and failure of a defective α-CP monolayer with a carbon divacancy strained along the AC direction. Since the puckers are oriented perpendicular to the direction of the applied tensile strain, the initial stage of the deformation is accomplished by flattening of the puckers. At a larger strain, the flattening mode switches to the bond stretching mode, after which, the PP bonds oriented along the AC direction are mostly strained, especially around an embedded vacancy.

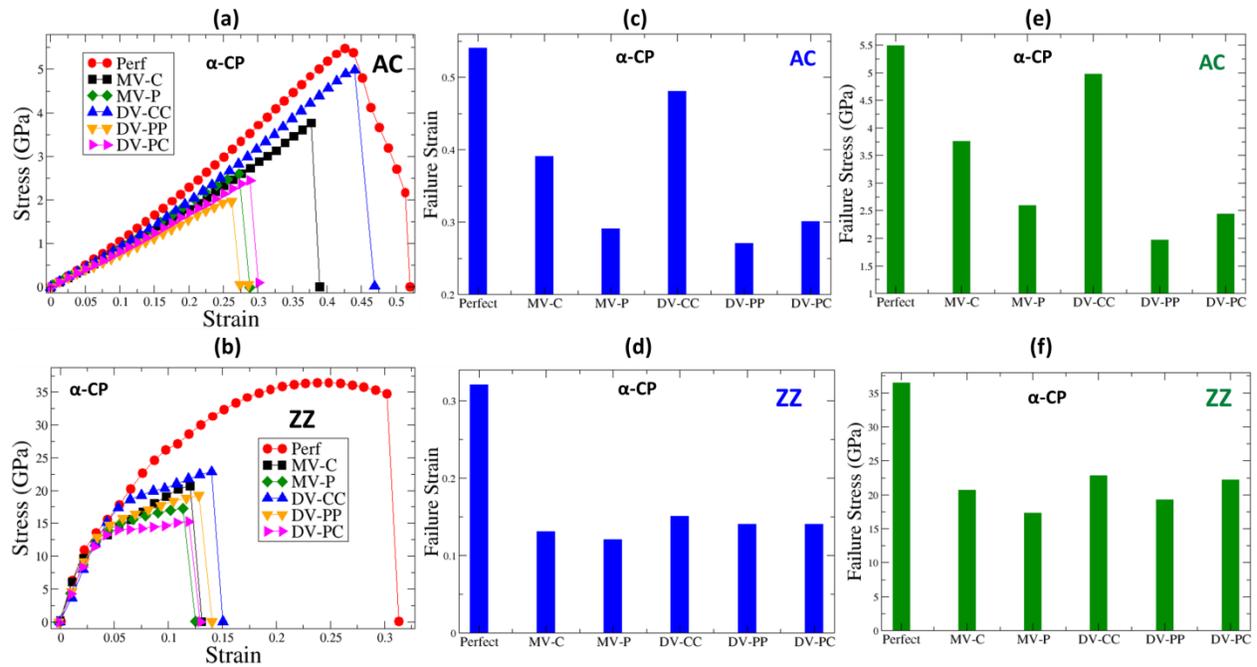

**Figure 8:** Stress-strain curves of pristine (red circles) and defective α-CP monolayers containing a monovacancy (black squares), phosphorous monovacancy (green diamonds), carbon divacancy (blue up triangles), phosphorous divacancy (yellow down triangles) and mixed phosphorus-carbon divacancy (magenta right triangles) obtained along the AC (a) and ZZ (b)

directions. The failure strain of pristine and defective α-CP calculated along the AC (c) and ZZ (d) directions, respectively. The failure stress of α-CP calculated along the AC (e) and ZZ (f) directions, correspondingly.

Since carbon vacancy breaks the uniformity of the sample (see Figure 10(a)), as a result, the atomistic configuration in the vicinity of the hosted vacancy is modified. The length of the bonds and the angles between them are adjusted accordingly around the divacancy. The length of several PP bonds located around the divacancy increases more rapidly than that of the rest under applied tensile strength.

In order to gain a better understanding of the vacancy effect, we selected a few pairs of bonded P-atoms located at various distances away from the hosted DV-CC divacancy (see the pairs of P-atoms specified by green, blue, red and black in Figure 10(a) at ε=0.45). The bond length of the selected pairs of atoms was measured as a function of applied strain (see Figure 10(b)). As can be seen in Figure 10(b), the lengths of the selected PP bonds increase insignificantly in the phase of flattening of the puckers via bond rotation (ε≲0.3). At the bond stretching phase (ε≳0.3), the bond lengths start to increase rapidly. Although the selected bonds are located at various distances from the carbon divacancy, the dependence of the bond lengths to the applied tensile strain is rather similar (see Figure 10(b)). As can be seen in Figure 10(b), the initial bond length (at ε=0) of the $P_0P_0$ bond belonging to one of the pentagons of the MV-CC (5|8|5) divacancy (see red circles in Figure 10(a)) is the smallest one, but the rate at which it increases with applied tensile strain is similar to others (see Figure 10(b)). As the tensile strain continues to increase, the bond length of the PP bonds located around the divacancy (see the bond length of the $P_1P_1$ pair of P-atoms marked by black in Figure 10(a)) rises sharply and eventually exceeds the bond length limit (indicated by the dotted red line in Figure 10(b)) at a critical strain ε=0.47. It is worth noting that due to the presence of the strong CC bonds, the failure happens around the carbon divacancy, but not at the divacancy core: the $P_1P_1$ bond that fractures first belongs to one of the PP bond rows located close to the divacancy core (see Figure 10(a)). The initial fracture of the PP bonds in the vicinity of the hosted divacancy induces the subsequent failure of the neighboring PP bonds. A crack, which initiates at the vacancy core and propagates perpendicular to the direction of applied tensile strain, causes the brittle-like fracture of α-CP monolayer along the AC direction (see Figure 10(b)).

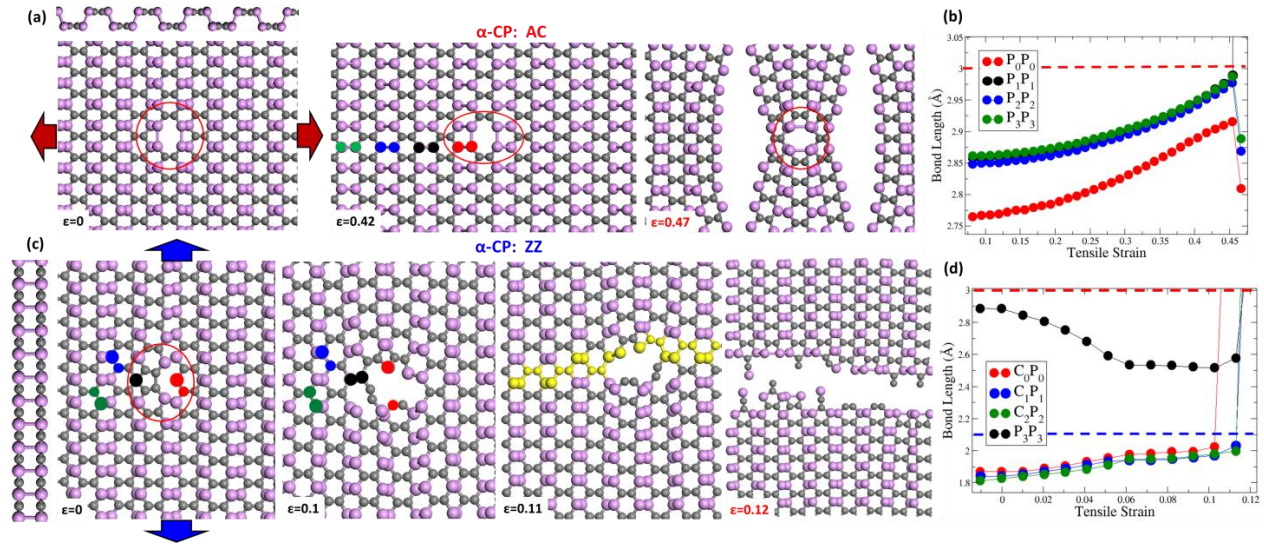

Figure 9: (a) Deformation and failure of α-CP monolayer with a carbon divacancy (DV-CC) under tensile strain applied along the AC direction. The atomistic images are taken at ε=0 (top and side view), ε=0.42 and $\varepsilon_f$=0.47 (top views). The arrows indicate the direction of the applied tensile strain. The red loop around the vacancy is to guide the eye. (b) Length of the phosphorous-phosphorous (PP) bonds connecting a selected group of atoms: $P_0P_0$ (red circles), $P_1P_1$ (black circles), $P_2P_2$ (blue

circles) and P₃P₃ (green circles) as a function of tensile strain. The designated atoms are indicated by the corresponding color in the atomistic image taken at ε=0.42 in (a). The red dotted line indicates the maximal length of PP bond. (c) Deformation and failure of α-CP monolayer with a phosphorous monovacancy (MV-P) strained along the ZZ direction. The atomistic images are taken at ε=0 (top and side view), ε=0.1, ε=0.11 and $\varepsilon_f$=0.12 (top views). The bonds between the highlighted atoms (at ε=0.11) rupture at the failure strain, forming the crack pathway. (d) Length of the carbon-phosphorous (CP) and PP bonds connecting a selected group atoms: C₀P₀ (red circles), C₁P₁ (blue circles) and C₂P₂ (green circles) and P₃P₃ (black circles), as a function of tensile strain. The chosen atoms are indicated by the respective color in the atomistic image taken at ε=0.1 in (c). The red and blue dotted lines indicate the maximal lengths of PP and CP bonds, correspondingly.

## *Failure along the ZZ direction*

The deformation and failure of defective α-CP containing a phosphorous monovacancy uniaxially stretched along the ZZ direction is shown in Figure 10(c). In contrast to the previous case, the direction of applied tensile strain coincides with the orientation of the puckers. Therefore, the sample is deformed via direct stretching of the CP bonds oriented at an acute angle to the ZZ direction. Yet, the presence of a single vacancy disrupts the uniformity of the tensile deformation and changes the failure mechanism of α-CP monolayer stretched along the ZZ direction. To gain a better insight into the impact of the hosted MV-P monovacancy on the deformation and the failure, a few pairs of bonded C- and P-atoms located around the vacancy were selected: a pair C₀P₀ of bonded atoms belonging to one of the monovacancy rings (see the red atoms in Figure 10(c)), and two pairs of C₁P₁ and C₂P₂ of atoms located in the vacancy vicinity (see the blue and the green atoms in Figure 10(c)). In addition, a pair P₃P₃ of bonded P-atoms was selected for comparison: one of the P-atoms is in the pentagon ring of the MV-P divacancy (see the black atoms in Figure 10(c)).

The bond lengths of the selected pairs of atoms were measured as a function of applied tensile strain (see Figure 10(d)). Initially, the bond lengths of the selected CP bonds increase gradually with applied tensile strain. Among the selected bonds, the length of the C₀P₀ bond is initially (at ε=0) slightly larger, and increases slightly faster with applied tensile strain. At tensile strain ε=0.1, it reaches the bond length limit and ruptures (see Figure 10(c, d)). Simultaneously rupture also occurs for the CC bond in the pentagon ring of the MV-P monovacancy oriented along the direction of applied strain and a few CP bonds around the MV-P monovacancy. Hence, the failure of the defective α-CP monolayer is initiated in the vacancy core, and then spreads around with increasing tensile strain. Ultimately, at a critical strain $\varepsilon_{cr}^{ZZ}$=0.12, the stretched α-CP monolayer undergoes a brittle-like fracture: it breaks into two halves (see Figure 10(c) at ε=0.12) when the CP bonds in the vacancy vicinity rupture (see the sharp increase in the length of the C₁P₁ and C₂P₂ bonds in Figure 10(d)). The crack is initiated at the vacancy core, and then propagates across the defected monolayer perpendicular to the direction of the applied strain. We reconstructed the crack path by tracing back the atoms whose bonds were broken at the critical strain (see the highlighted atoms at ε=0.11 in Figure 10(c)). Although the majority of the broken bonds are the CP bonds, there is a small fraction (~10%) of the ruptured PP bonds. For instance, the selected P₃P₃ bond was broken at the critical strain (see Figure 10(d)).

## *Failure strain and stress*

The obtained values of the failure strain and stress for defective α-CP monolayer uniaxially strained along the AC and the ZZ directions are plotted in Figure 8(c-f) and listed in Table 3. Evidently, the failure strain and stress of defective α-CP monolayer are affected in a different way by various types of vacancies.

Along the AC direction, the failure strain is mainly affected by phosphorous vacancies (see Figure 8(c)): an introduction of a DV-PP divacancy reduces the failure strain $\varepsilon_{AC}$ by $\frac{\varepsilon_{AC}-\varepsilon_{AC}^p}{\varepsilon_{AC}^p} \cdot 100\% \approx$ -50% (where $\varepsilon_{AC}^p$ is the failure strain of pristine α-CP monolayer along the AC direction); while a MV-P monovacancy reduces it by $\frac{\Delta\varepsilon_{AC}}{\varepsilon_{AC}^p}$ ≈-46%. In contrast, the impact of carbon vacancies is relatively weak: a DV-CC divacancy reduces the failure strain only by $\frac{\Delta\varepsilon_{AC}}{\varepsilon_{AC}^p}$ ≈-13% and a MV-C monovacancy by $\frac{\Delta\varepsilon_{AC}}{\varepsilon_{AC}^p}$ ≈-28%. The effect of a mixed DV-PC divacancy is comparable to that of phosphorous vacancies: the failure strain is reduced by $\frac{\Delta\varepsilon_{AC}}{\varepsilon_{AC}^p}$ ≈-44%.

Along the ZZ direction, the failure strain of α-CP is almost equally affected by phosphorous and carbon vacancies (see Figure 8(d)): both DV-PP and DV-PC divacancies decrease the failure strain by $\frac{\Delta\varepsilon_{ZZ}}{\varepsilon_{ZZ}^p}$ ≈-56%, while a DV-CC divacancy by $\frac{\Delta\varepsilon_{ZZ}}{\varepsilon_{ZZ}^p}$ ≈-49%. Similarly, a MV-P monovacancy reduces the failure strain by $\frac{\Delta\varepsilon_{ZZ}}{\varepsilon_{ZZ}^p}$ ≈-59%, while a MV-C monovacancy by $\frac{\Delta\varepsilon_{ZZ}}{\varepsilon_{ZZ}^p}$ =-58%.

Table 3: Young's modulus, failure stress and failure strain of pristine and defective α-CP obtained along the AC and the ZZ directions.

| Defect | α-CP (AC) | | | α-CP (ZZ) | | |
|---|---|---|---|---|---|---|
| | Y(GPa) | $\sigma_f$(GPa) | $\varepsilon_f$ | Y(GPa) | $\sigma_f$(GPa) | $\varepsilon_f$ |
| Perfect | 35.33 | 5.49 | 0.54 | 343.04 | 36.47 | 0.32 |
| MV-C | 28.84 | 3.76 | 0.39 | 260.15 | 20.72 | 0.13 |
| MV-P | 28.03 | 2.59 | 0.29 | 224.81 | 17.29 | 0.12 |
| DV-CC | 31.37 | 4.97 | 0.47 | 273.26 | 22.85 | 0.15 |
| DV-PP | 27.96 | 1.97 | 0.27 | 233.02 | 19.26 | 0.14 |
| DV-PC | 28.52 | 2.44 | 0.3 | 234.32 | 22.61 | 0.14 |

The failure stress of a defective α-CP monolayer strained along the AC direction is also to a large degree affected by the phosphorous vacancies (see Figure 8(e)): a DV-PP divacancy reduces the failure stress ($\sigma_{AC}$) by $\frac{\sigma_{AC}-\sigma_{AC}^p}{\sigma_{AC}^p} \cdot 100\%$ ≈-64% (where $\sigma_{AC}^p$ is the failure stress of pristine α-CP monolayer along the AC direction), while a MV-P monovacancy reduces it by $\frac{\Delta\sigma_{AC}}{\sigma_{AC}^p}$ ≈-53%. In contrast, the effect of carbon vacancies is noticeably weaker: a DV-CC divacancy decreases the failure strain only by $\frac{\Delta\sigma_{AC}}{\sigma_{AC}^p}$ ≈ -10%, whereas a MV-C monovacancy by $\frac{\Delta\sigma_{AC}}{\sigma_{AC}^p}$ ≈-32%. A mixed DV-PC divacancy, like a phosphorous vacancy, also reduces the failure stress considerably by $\frac{\Delta\sigma_{AC}}{\sigma_{AC}^p}$ ≈-55%.

The failure stress of defective α-CP monolayer stretched along the ZZ direction is mostly affected by the hosted phosphorous vacancies (see Figure 8(f)): a DV-PP divacancy reduces it by $\frac{\Delta\sigma_{ZZ}}{\sigma_{ZZ}^p}$ ≈-47% and a MV-P monovacancy by $\frac{\Delta\sigma_{ZZ}}{\sigma_{ZZ}^p}$ ≈-53%. The impact of carbon vacancies on the failure stress along the ZZ direction is comparable to that of phosphorous vacancies: a DV-CC divacancy decreases the failure strain by $\frac{\Delta\sigma_{ZZ}}{\sigma_{ZZ}^p}$ ≈-37%, while a MV-C monovacancy by $\frac{\Delta\sigma_{ZZ}}{\sigma_{ZZ}^p}$ ≈-43%. The effect of a mixed DV-PC divacancy is also substantial since the failure stress is reduced by $\frac{\Delta\sigma_{ZZ}}{\sigma_{ZZ}^p}$ ≈-38%.

## Deformation and failure of defective β-CP monolayers

Finally, we studied the deformation and failure of a defective β-CP monolayer stretched along the primary directions. The obtained stress-strain curves for the defective monolayer uniaxially strained along the AC and ZZ directions are shown in Figure 10(a, b), respectively. As can be in Figure 10 (a, b), the hosted vacancies significantly reduce both the failure strain and the failure stress of β-CP. Clearly, different types of vacancies affect the failure strain and the failure stress differently. Along the AC direction, the failure strain and the failure stress are mainly affected by phosphorous vacancies, while the impact of carbon vacancies is relatively weaker. Along the ZZ direction, the failure strain and the failure stress are affected in the same way by both phosphorous and carbon vacancies.

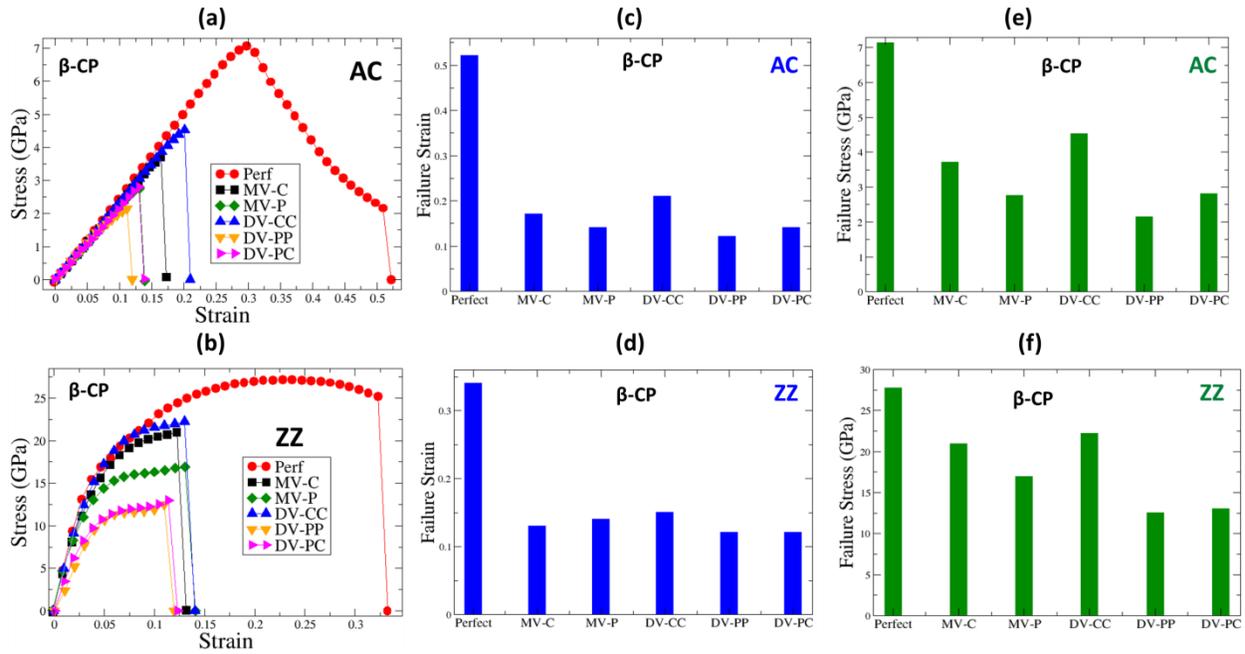

Figure 10: Stress-strain curves of pristine (red circles) and defective β-CP monolayers containing carbon monovacancy (black squares), phosphorous monovacancy (green diamonds), carbon divacancy (blue up triangles), phosphorous divacancy (yellow down triangles) and mixed phosphorus-carbon divacancy (magenta right triangles) obtained along the AC (a) and the ZZ (b) directions. The failure strain of pristine and defective β-CP calculated along the AC (c) and the ZZ (d) directions, respectively. The failure stress of β-CP calculated along the AC (e) and the ZZ (f) directions, correspondingly

### *Failure along the AC direction*

The tensile deformation and the subsequent failure of a β-CP monolayer containing a phosphorous DV-PP divacancy stretched along the AC direction are illustrated in Figure 11(a). Like in the preceding case, when a defective α-CP monolayer was stretched along the AC direction, the overextended PP bonds around the hosted vacancy play the crucial role in the deformation and the ensuing failure of the β-CP monolayer.

In the vicinity of the DV-PP divacancy, the bond length and the bond angles of the surrounding atoms are adjusted to incorporate the defect. In order to examine the deformation and failure mechanism in detail, a few pairs of bonded P-atoms located in the neighborhood of the hosted divacancy (see the $P_0P_0$ pair of red atoms and the $P_1P_1$ pair of blue atoms in Figure 11(a)) were selected. For comparison, the $P_2P_2$ pair located at some distance away from the divacancy was chosen (see the $P_2P_2$ pair of black atoms in Figure 11(a)). The bond lengths of the selected PP bonds as a function of applied tensile strain are shown in Figure 11(b).

In the reference configuration (at ε=0), the initial bond lengths of the $P_0P_0$ and $P_1P_1$ bonds located in the vacancy neighborhood are smaller than that of the $P_2P_2$ bond situated at a larger distance (see Figure 11(b)). With increasing the applied tensile strain, the bond length of the $P_0P_0$ bond increases to the largest extent. It rapidly exceeds the lengths of other bonds and approaches the bond length limit first. At tensile strain ε=0.09, the $P_0P_0$ bond ruptures. Eventually, at a critical strain $\varepsilon_{cr}^{AC}$=0.14, the entire array of the PP bonds adjacent to the divacancy fractures (see Figure 11(a)). The brittle-like fracture, which is initiated at the vacancy neighborhood, proceeds via PP bond breaking as illustrated in Figure 11(a). The corresponding crack propagates normal to the direction of the applied tensile strain.

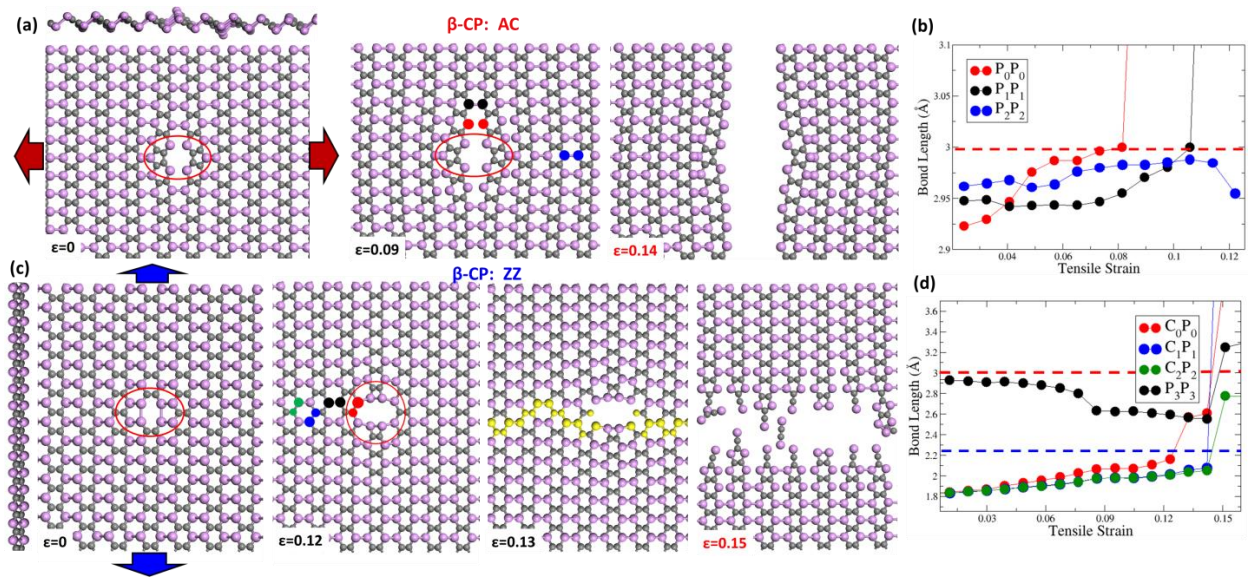

Figure 11: (a) Deformation and failure of β-CP monolayer with a phosphorous divacancy (DV-PP) under tensile strain applied along the AC direction. The atomistic images are taken at ε=0 (top and side view), ε=0.09 and $\varepsilon_f$=0.14 (top views). The arrows indicate the direction of the applied tensile strain. The red loop around the vacancy is to guide the eye. (b) Length of the PP bonds connecting a selected group of atoms: $P_0P_0$ (red circles), $P_1P_1$ (black circles) and $P_2P_2$ (blue circles) as a function of tensile strain. The chosen atoms are indicated by the corresponding color in the image taken at ε=0.09 in (a). The red dotted line indicates the maximal length of PP bond. (c) Deformation and failure of β-CP monolayer with a carbon divacancy (DV-CC) strained along the ZZ direction. The atomistic images are taken at ε=0 (top and side view), ε=0.12, ε=0.13 and $\varepsilon_f$=0.15 (top views). The bonds between the highlighted atoms (at ε=0.13) rupture at the failure strain, forming the crack pathway. (d) Length of the bonds connecting a selected group atoms: $C_0P_0$ (red circles), $C_2P_2$ (blue circles), $C_3P_3$ (green circles) and $P_3P_3$ (black circles), as a function of tensile strain. The selected atoms are indicated by the respective color in the atomistic image taken at ε=0.12 in (c). The blue and red dotted lines indicate the maximal length of CP and PP bonds, respectively.

### *Failure along the ZZ direction*

Lastly, the deformation and failure mechanism of a defective β-CP monolayer subjected to uniaxial tensile strain along the ZZ direction is demonstrated with a carbon DV-CC divacancy. It was found that the hosted carbon divacancy undergoes structural transformation under applied tensile strain: the two PP bonds shared by the adjacent pentagon-octagon-pentagon rings (and aligned in parallel to the direction of applied tensile strain) rupture at ε=0.12 as shown in Figure 10(c). At the same time, two new PP bonds (oriented perpendicular to the direction of applied strain) are formed between the nearest neighboring P-atoms of the adjacent pentagon rings since the distance between the P-atoms decreases sufficiently due to transverse contraction of β-CP. As a result, the original (5|8|5) DV-CC divacancy is transformed into an extended (4|10|4) divacancy (see Figure 11 (c) at ε=0.12).

Under uniaxial tensile strain applied along the ZZ direction, the highest load is borne by the CP bonds since they oriented at an acute angle to the direction of applied strain, while the CC and PP bonds are oriented normal to it. The applied tensile strain elongates the CP bonds until they fracture at a critical strain. In pristine β-CP monolayer, all the CP bonds are stretched to the same extent. While in a defective sample, this uniformity of the bond deformation is broken around the hosted vacancy. Several CP bonds located in the vicinity of the hosted DV-CC divacancy were selected to examine in detail the effect of the embedded divacancy on the deformation and the failure mechanism. The following three CP bonds were chosen: the $C_0P_0$ bond of the decagon ring of the (4|10|4) divacancy (see red atoms in Figure 11(c)) and $C_1P_1$ and $C_2P_2$ bonds located nearby (see the blue and green atoms in Figure 11(c), correspondingly). For comparison, the $P_3P_3$ bond formed by a pair of P-atoms adjacent to the DV-CC divacancy (see a pair of black atoms in Figure 11(c)) was taken. The lengths of the selected bonds as a function of applied tensile strain are shown in Figure 11(d). The bond lengths of the CP bonds gradually increase with applied tensile strain, while the length of the $P_3P_3$ bond decreases due to contraction of the β-CP monolayer in the transverse direction. The length of the $C_0P_0$ bond increases faster than the lengths of other bonds (see Figure 11(d)). Correspondingly, the first bond that ruptures in the vicinity of the carbon divacancy is the $C_0P_0$ bond. As can be seen in Figure 11(c), there are four equivalent $C_0P_0$-type bonds rupturing at ε=0.13.

The fracture of the CP bonds at the divacancy core leads to the subsequent fracture of their adjacent CP bonds with increasing tensile strain. At a critical strain $\varepsilon_{cr}^{ZZ}$=0.15, the defective β-CP monolayer undergoes a brittle-like failure (see Figure 11(c)). The crack path formed by the ruptured bonds can be reconstructed by identifying the atoms whose bonds were broken at the critical strain (see the highlighted atoms at ε=0.13 in Figure 11(c)). Analysis of the atomistic configuration of the fractured β-CP monolayer indicates that in addition to the CP bonds, there is a small fraction (~10%) of the PP bonds rupturing at the failure strain. For example, the selected $P_3P_3$ bond breaks at the failure strain as can be seen in Figure 11(c, d).

### *Failure strain and stress*
The effect of single and double vacancies on the failure strain and stress of defective β-CP monolayer stretched along the AC and ZZ directions is illustrated in Figure 10(c, e) and in Figure 10(d, f), correspondingly. The calculated values of the failure strain and stress are reported in Table 4.

As can be seen in Figure 10(c), the failure strain of defective β-CP monolayer stretched along the AC direction is mostly affected by phosphorous vacancies: a DV-PP divacancy reduces the failure strain by $\frac{\Delta\varepsilon_{AC}}{\varepsilon_{AC}^p} \approx$-77%, while a MV-P monovacancy by $\frac{\Delta\varepsilon_{AC}}{\varepsilon_{AC}^p} \approx$-73%. The effect of carbon vacancies is only slightly weaker: a DV-CC divacancy reduces the failure strain by $\frac{\Delta\varepsilon_{AC}}{\varepsilon_{AC}^p} \approx$-60%, yet a MV-C monovacancy decreases it by $\frac{\Delta\varepsilon_{AC}}{\varepsilon_{AC}^p} \approx$-67%. The impact of a mixed carbon-phosphorous divacancy is also significant: a DV-CP divacancy reduces the failure strain by $\frac{\Delta\varepsilon_{AC}}{\varepsilon_{AC}^p} \approx$-73%.

When a defective β-CP monolayer stretched along the ZZ direction, the failure strain is particularly sensitive to phosphorous vacancies (see Figure 10(d)): a DV-PP divacancy reduces the failure strain by $\frac{\Delta\varepsilon_{ZZ}}{\varepsilon_{ZZ}^p} \approx$-65%, while a MV-P monovacancy by $\frac{\Delta\varepsilon_{ZZ}}{\varepsilon_{ZZ}^p} \approx$-59%. Along the ZZ direction, the impact of the carbon vacancies is comparable to phosphorous ones: a DV-CC divacancy decreases the failure strain by $\frac{\Delta \varepsilon_{ZZ}}{\varepsilon_{ZZ}^p} \approx$ -56%, whereas a MV-C monovacancy by $\frac{\Delta \varepsilon_{ZZ}}{\varepsilon_{ZZ}^p} \approx$ -62%. A mixed DV-PC divacancy reduces the failure strain by $\frac{\Delta \varepsilon_{ZZ}}{\varepsilon_{ZZ}^p} \approx$ -62%.

Table 4: Young's modulus, failure stress and failure strain of pristine and defective β-CP obtained along the AC and the ZZ directions.

| Defect | β-CP (AC) | | | β-CP (ZZ) | | |
|---|---|---|---|---|---|---|
| | $Y$(GPa) | $\sigma_f$(GPa) | $\varepsilon_f$ | $Y$(GPa) | $\sigma_f$(GPa) | $\varepsilon_f$ |
| Perfect | 81.25 | 7.15 | 0.52 | 288.85 | 27.25 | 0.34 |
| MV-C | 72.67 | 3.71 | 0.17 | 239.11 | 20.98 | 0.13 |
| MV-P | 70.01 | 2.76 | 0.14 | 218.88 | 16.96 | 0.14 |
| DV-CC | 73.61 | 4.53 | 0.21 | 251.58 | 22.24 | 0.15 |
| DV-PP | 65.71 | 2.15 | 0.12 | 159.33 | 12.52 | 0.12 |
| DV-PC | 70.43 | 2.81 | 0.14 | 177.57 | 13.01 | 0.13 |

The failure stress of defective β-CP monolayer strained along the AC direction is also mainly affected by phosphorous vacancies (see Figure 10(e)): a DV-PP divacancy reduces the failure stress by $\frac{\Delta \sigma_{AC}}{\sigma_{AC}^p} \approx$ -70%, while a MV-P monovacancy by $\frac{\Delta \sigma_{AC}}{\sigma_{AC}^p} \approx$ -61%. The impact of carbon vacancies is relatively moderate: a DV-CC divacancy reduces the failure stress only by $\frac{\Delta \sigma_{AC}}{\sigma_{AC}^p} \approx -37\%$, while a MV-C monovacancy by $\frac{\Delta \sigma_{AC}}{\sigma_{AC}^p} \approx$ -48%. A mixed DV-PC divacancy reduces the failure stress by $\frac{\Delta \sigma_{AC}}{\sigma_{AC}^p} \approx$ -61%.

We note that the failure stress defective β-CP monolayer stretched along the ZZ direction is particularly sensitive to phosphorous vacancies (see Figure 10(f)): a DV-PP divacancy reduces the failure stress by $\frac{\Delta \sigma_{ZZ}}{\sigma_{ZZ}^p} \approx$ -54%, while a MV-P monovacancy by $\frac{\Delta \sigma_{ZZ}}{\sigma_{ZZ}^p} \approx$ -38%. In contrast, the impact of carbon vacancies is comparatively weaker than that of phosphorous vacancies: a DV-CC divacancy decreases the failure stress by $\frac{\Delta \sigma_{ZZ}}{\sigma_{ZZ}^p} \approx$ -18%, whereas a MV-C monovacancy by $\frac{\Delta \sigma_{ZZ}}{\sigma_{ZZ}^p} \approx$ -23%. A mixed DV-PC divacancy decreases the failure strain by $\frac{\Delta \sigma_{ZZ}}{\sigma_{ZZ}^p} \approx$ -52%.

## 4. Conclusions

Using the DFTB method, we investigated the elastic properties, deformation and failure behaviors of pristine and defective CP monolayers subjected to uniaxial tensile strain along the primary directions (AC and ZZ). Two variants of carbon-phosphide (α-CP and β-CP) were studied and two types of vacancies (monovacancies and divacancies) were considered.

We examined the atomistic configurations around the hosted vacancies and calculated the corresponding vacancy formation energies. It was found that the removal of carbon atoms leads to a (5|9) configuration for a carbon monovacancy and a (5|8|5) configuration for a carbon divacancy. However, the deletion of phosphorous atoms significantly disturbs the local lattice structure in the vicinity of the embedded vacancy: the removal of a P-atom leads to the formation of an extended (5|17) phosphorous monovacancy, while the elimination of a pair of adjacent P-atoms results in an expanded

(5|16|5) divacancy. The removal of a couple of neighboring P- and C-atoms leads to an extended (5|11) mixed divacancy. The formation energy of carbon divacancy is the lowest one, followed by the slightly higher value for the carbon monovacancy in both α and β carbon-phosphide monolayers. In contrast, the formation energy of a phosphorous divacancy is the highest one.

We measured the Young's modulus and the in-plane and out-of-plane Poisson's ratios of pristine CP monolayers along the AC and ZZ directions. A strong mechanical anisotropy was found to be characteristic for both allotropes: they are particularly stiff in the ZZ direction and comparatively soft in the AC direction: the Young's modulus along the ZZ direction is by one order of magnitude larger than that along the AC direction. Furthermore, the Young's modulus of CP allotropes along the ZZ direction exceeds that of phosphorene, whereas along the AC direction, they are comparable. The calculated Poisson's ratios also exhibit a strong anisotropy: the in-plane Poisson's ratios along the primary directions differ by two orders of magnitude.

The effects of the hosted vacancies on the elastic properties of CP monolayers were investigated. The Young's modulus is strongly affected by phosphorous vacancies: it is reduced up to $\Delta Y/Y_p \approx -34\%$ by a P-monovacancy in α-CP, and up to $\Delta Y/Y_p \approx -45\%$ by a P-divacancy in β-CP. The effect of carbon vacancies is relatively weaker: the Young's modulus is reduced up to $\Delta Y/Y_p \approx -24\%$ by a C-monovacancy in α-CP, and up to $\Delta Y/Y_p \approx -17\%$ a C-monovacancy in β-CP. The Young's modulus along the AC direction is more sensitive to the hosted vacancies then along the ZZ direction in both CP allotropes. The in-plane and out-plane Poisson's ratios are only slightly increased by vacancies.

The tensile deformation and failure mechanism of pristine of α- and β-CP monolayers were examined. Due to the puckered structure of CP, the deformation and failure mechanisms depend on the direction of applied tensile strain. At relatively low tensile strain, the PP bonds oriented along the AC direction rotate at the hinge-like structures of the puckers, and as a result, the puckered structure is flattened. At high stain, the weakest PP bonds are strained most. When they rupture at a critical strain, a CP monolayer loses its structural stability and fails. Since the deformation includes the flattening of the puckered structure via the bond rotation, the α- and β-CP monolayers can be stretched up to $\varepsilon_{AC}^p$=0.54 (α-CP) and $\varepsilon_{AC}^p$=0.52 (β-CP), respectively, which are comparable to pristine phosphorene $\varepsilon_{AC}^p$=0.5. The fracture strains along the AC direction for the two allotropes, $\sigma_{AC}^p$=5.49 GPa (α-CP) and $\sigma_{AC}^p$=7.15 GPa (β-CP), are also similar to that of phosphorene, $\sigma_{AC}^p$=8 GPa [43,50].

When a CP monolayer is stretched along the ZZ direction, the tensile deformation directly proceeds via stretching of the CP bonds, inclined along the ZZ direction. The strained CP bonds simultaneously rupture at a critical strain throughout the sample. The structural stability is lost due to bond fragmentation. Since CP bonds are stronger than PP bonds, the fracture strains along the ZZ direction: $\varepsilon_{ZZ}^p$=0.32 (α-CP) and $\varepsilon_{ZZ}^p$=0.34 (β-CP) exceed that of pristine phosphorene $\varepsilon_{ZZ}^p$=0.24 [43,50]. The fracture stresses along the ZZ direction: $\sigma_{ZZ}^p$=36.5GPa (α-CP) and $\sigma_{ZZ}^p$=27.5GPa (β-CP) are also larger than that of pristine phosphorene $\sigma_{ZZ}^p$=18GPa [43,50]. Pristine CP demonstrates a strong anisotropy, similar to phosphorene: the failure strain along the AC direction is almost twice as large as along the ZZ direction, while the failure stress along the ZZ direction is by an order of magnitude larger than that along the AC one.

We investigated the deformation and failure behaviors of defective α- and β-CP monolayers. A hosted vacancy breaks the deformation uniformity of the sample: a significant fraction of bonds is strained around it. Under tensile strain applied along the AC direction, the lengths of PP bonds oriented along the direction of applied tensile strain are most increased around the vacancy site. At a critical strain, these PP bonds rupture first, inducing the subsequent failure of the adjacent PP bonds. The resultant crack leads to a brittle-like failure of the defective monolayer. Similarly, when the defective monolayer is stretched along the ZZ direction, the lengths of the CP bonds oriented along to the direction of applied strain grow faster in the vicinity of a hosted vacancy. These CP bonds fracture first at the outset of the brittle-like failure. The resultant crack, which is nucleated around the vacancy, moves perpendicular to the direction of the applied tensile strain, and break the adjacent CP bonds along its way. The failure strain and stress along the AC direction are mainly affected by phosphorous vacancies, while the impact of carbon vacancies is relatively weaker. Along the ZZ direction, both the failure strain and stress are almost equally affected by phosphorous and carbon vacancies.

The examined pristine and defective α-CP and β-CP monolayers can be potentially useful for applications in nanoelectromechanical systems. Since the presence of point defects cannot be avoided in general, it is important to understand their effects on the mechanical properties. Such understanding can help us set specific restrictions on the mechanical strains and stresses for CP monolayers used in nanoelectromechanical systems. We expect that the present findings may provide useful guidelines for the design, fabrication and applications of α and β CP monolayers.

## 5. Acknowledgements

This work was supported by the A*STAR Computational Resource Centre through the use of its high performance computing facilities, and by a grant from the Science and Engineering Research Council (152-70-00017).

## Citations